\documentclass[referee]{cjaa}           

\usepackage{graphicx}                   
\input{epsf.sty}                        
\input{psfig.sty}                       

\begin{document}

   \title{Thermal Bremsstrahlung Radiation in a Two-Temperature
   Plasma}

   \volnopage{Vol.0 (200x) No.0, 000--000}      
   \setcounter{page}{1}          

   \author{Bin LUO
      \inst{1,2}\mailto{}
   \and Shuang Nan Zhang
      \inst{1,2}
}

\offprints{B. LUO}                   

   \institute{Physics Department and Center for Astrophysics, Tsinghua
             University, Beijing 100084, China\\
             \email{luobin@tsinghua.org.cn}
        \and
             Institute of High Energy Physics, Chinese Academy of
sciences, Beijing 100039, China}

   \date{Received~~2003 month day; accepted~~2003~~month day}

   \abstract{
   In the normal one-temperature plasma the motion of ions is usually
neglected when calculating the Bremsstrahlung radiation of the
plasma. Here we calculate the Bremsstrahlung radiation of a
two-temperature plasma by taking into account of the motion of
ions. Our results show that the total radiation power is always
lower if the motion of ions is considered. We also apply the
two-temperature Bremsstrahlung radiation mechanism for an
analytical Advection-Dominated Accretion Flow (ADAF) model; we
find the two-temperature correction to the total Bremsstrahlung
radiation for ADAF is negligible.
   }

   \authorrunning{B. LUO \& S. Nan ZHANG}            
  \titlerunning{Thermal Bremsstrahlung Radiation in a Two-Temperature
   Plasma}  

   \maketitle

%
%
\section{Introduction}           
\label{sect:intro}
In a plasma, electrons are constantly accelerated during their
collisions with ions, leading to Bremsstrahlung radiation.
Usually, when calculating the radiation power of a plasma, the
motion of ions is neglected, because the ion's mass is much higher
than the electron mass. However the motion of ions may not be
neglected if the ion temperature is much higher than that of
electrons, such as in the ADAF model, in which the temperature of
ions $\sim10^{11}$\,k may be much higher than that of electrons
($10^8\sim10^9$\,k) (Narayan \& Yi, 1995). Under such condition,
the velocities of ions are comparable to or even higher than that
of electrons. Therefore the motion of ions must be taken into
account for calculating the thermal Bremsstrahlung radiation of a
two-temperature plasma.

Our calculations show that the radiation
power is reduced significantly if the ion temperature is much
higher than the electron temperature. However when applying the
two-temperature Bremsstrahlung radiation to the ADAF model, we find the
total Bremsstrahlung radiation emissivity is not significantly different
from the one-temperature Bremsstrahlung radiation, because most of
the Bremsstrahlung radiation is produced at large radii where the ion's
temperature is not significantly different from the electron temperature.\\



\section{Thermal Bremsstrahlung Radiation}
\label{sect:TBR}
We begin by considering an individual scattering event between an
electron and an ion in which an electron with a velocity $v$ and
an impact parameter $b$ is scattered by an ion; we assume the
electron's motion is non-relativistic. Then the radiation power
emitted at a specific frequency by this electron is given by
(Padmanabhan, 2000, p. 295)
\begin{equation}
W_1(\omega)=\frac{8Z^2 e^6}{3\pi c^3 m_e^2 v^2 b^2}.
\end{equation}
Integrating over $b$ with the limit $b_{min}=Ze^2/m_e v^2$ and
$b_{max}=v/\omega$ (YOU, 1998, p. 282), we get
\begin{equation}
P_1(\nu)=2\pi P_1(\omega)=4\pi^2 N_z v
\int_{b_{min}}^{b_{max}}W_1(\omega)b \mathrm{d} b=\frac{32\pi N_Z
Z^2 e^6}{3c^3 m_e^2 v}\ln\frac{m_e v^3}{2\pi Z e^2\nu},
\end{equation}
where $N_Z$ is the number density of ions. Integrating it over all
electrons and assuming the electrons follow the non-relativistic
Maxwellian velocity distribution, we get the specific emissivity,
\begin{equation}
j_1(\nu)=\frac{128\pi^2 Z^2 e^6}{3c^3 m_e^2}N_ZN_e(\frac{m_e}{2\pi
kT})^{3/2}\int_0^\infty \exp (-\frac{m_ev^2}{2kT})v\ln\frac{m_e
v^3}{2\pi Ze^2\nu} \mathrm{d} v,
\end{equation}
where $N_e$ is the number density of electrons.

\section{Two-Temperature Bremsstrahlung Radiation}
\label{sect:TTBR}
Assuming that the temperature of ions and electrons are $T_Z$ and
$T_e$ respectively, we still begin by considering an individual
scattering event. The radiation power of an electron in the
rest-frame of an ion is similar to Eq.~(1),
\begin{equation}
W_2(\omega)=\frac{8Z^2e^6}{3\pi c^3 m_e^2 v^2 b^2},
\end{equation}
except that $v$ is the relative velocity between the ion and the
electron
\begin{displaymath}
v=(v_e^2+v_Z^2-2v_ev_Z\cos\theta)^{1/2}.
\end{displaymath}
Integrating over $b$, we get
\begin{equation}
P_2(\nu)=2\pi P_2(\omega)=4\pi^2\int_0^\infty \mathrm{d} \vec{v}_Z
\int_{b_{min}}^{b_{max}}W_2(\omega)N(\vec{v}_Z)v b\mathrm{d} b.
\end{equation}
Assuming that the ions and electrons all follow the non-relativistic Maxwellian
velocity distribution, and integrating $P_2(\omega)$ over all the
electrons, we get the specific emissivity,

\begin{eqnarray} j_2(\nu)=\frac{256\pi^3Z^2e^6}{3c^3m_e^2}N_ZN_e
(\frac{m_e}{2\pi kT_e})^{3/2}(\frac{m_Z}{2\pi
kT_Z})^{3/2}\int_0^\infty \exp(-\frac{m_ev_e^2}{2kT_e})v_e^2
\mathrm{d}v_e \nonumber\\\int_0^\infty
\exp(-\frac{m_Zv_Z^2}{2kT_Z})v_Z^2 \mathrm{d}v_Z
\int_0^\pi\ln\frac{m_ev^3}{2\pi Ze^2\nu}\frac{1}{v}\sin\theta
\mathrm{d}\theta.\end{eqnarray}
Because we are only considering the
non-relativistic regime, the radiation power observed in the
rest-frame of an ion is the same as that in the laboratory frame.

\section{Results}
\label{sect:Resu}
We make numerical calculations of the two-temperature plasma
radiation; all ions are assumed to be protons. The results are
shown in Figs.~1--3. We can see that in all cases the
two-temperature Bremsstrahlung radiation emissivity
is lower than the one-temperature case, and the difference is greater for higher electron and/or ion temperatures.\\

\begin{figure}[h]
  \begin{minipage}[t]{0.5\linewidth}
  \centering
  \includegraphics[width=65mm,height=50mm]{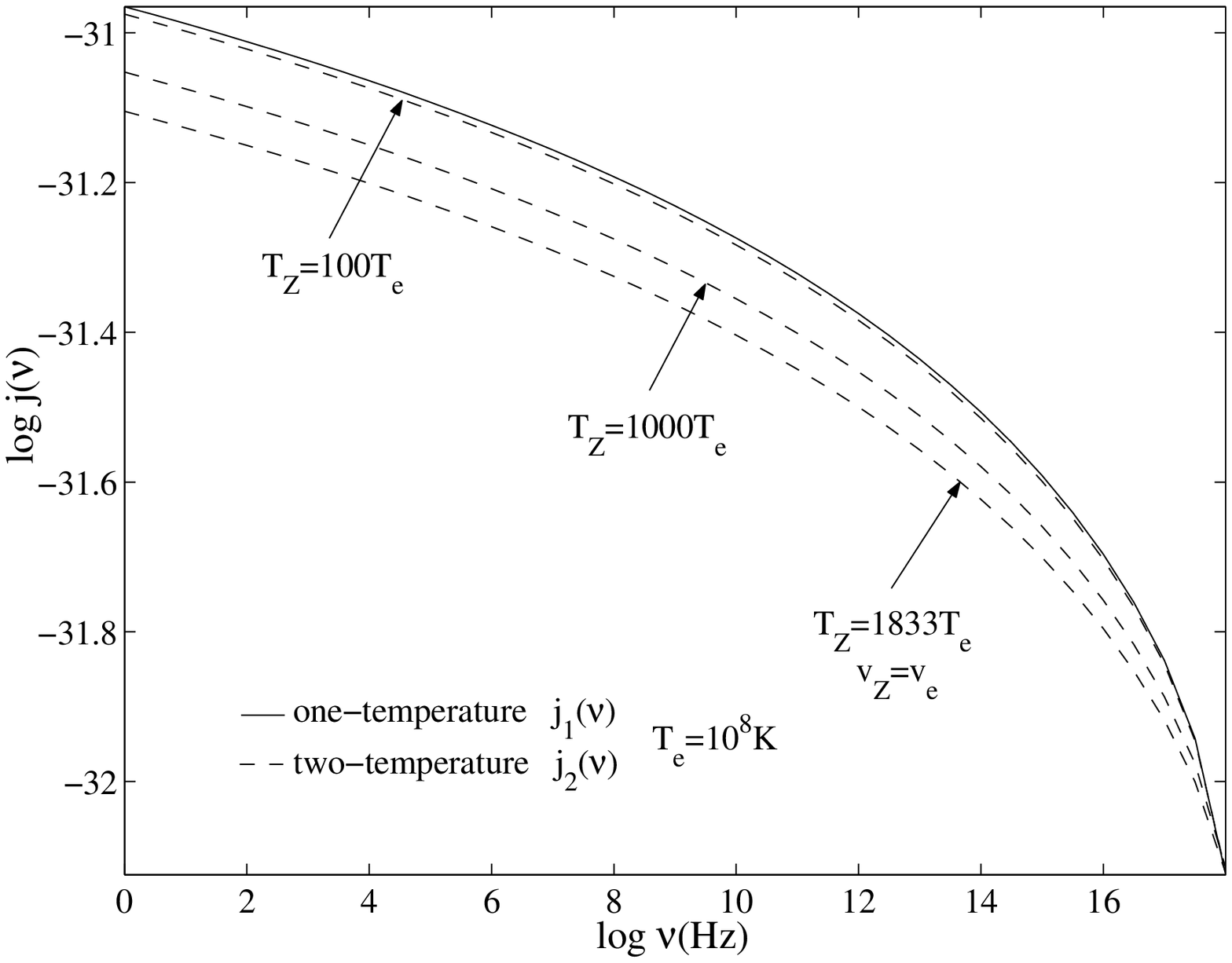}
  \vspace{-5mm}
  \caption{The specific emissivity of the
one-temperature and two-temperature Bremsstrahlung radiation.
$T_e=10^8$\,k. If $T_Z=1833T_e$, corresponding to the same average
velocities of ions and electrons, the radiation power of the
two-temperature Bremsstrahlung radiation is about 75\% of the
one-temperature Bremsstrahlung radiation. }
  \end{minipage}%
  \begin{minipage}[t]{0.5\textwidth}
  \centering
  \includegraphics[width=65mm,height=50mm]{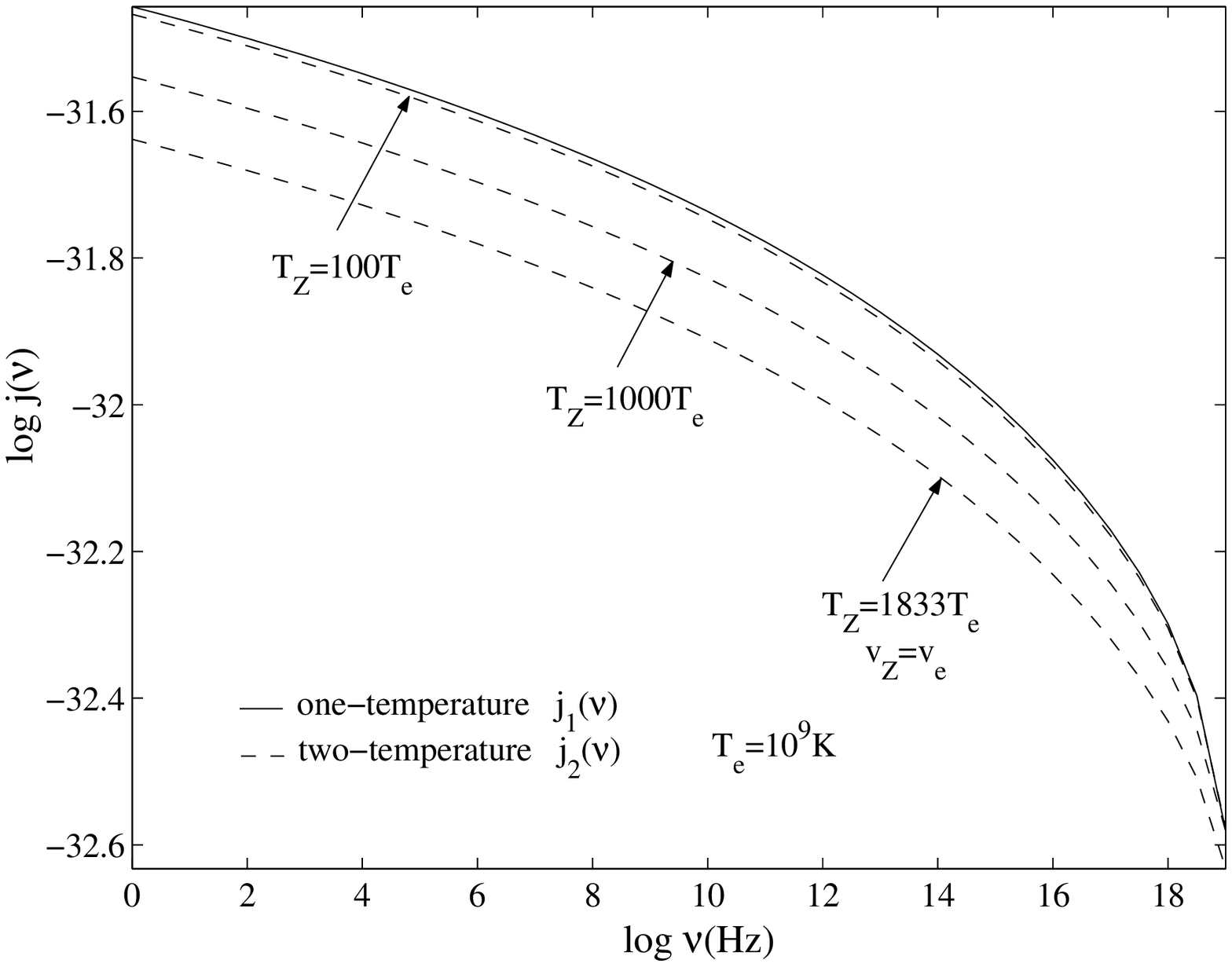}
  \vspace{-5mm}
  \caption{ The specific emissivity of the
one-temperature and two-temperature Bremsstrahlung radiation.
$T_e=10^9$\,k. If $T_Z=1833T_e$, the radiation power of the
two-temperature Bremsstrahlung radiation is about 68\% of the
one-temperature Bremsstrahlung radiation.}
  \end{minipage}%
  \label{Fig:fig12}
\end{figure}

We then calculate the total luminosity following an analytical
ADAF model (Mahadevan, 1997), assuming spherical accretion and
with all the self-similar equations as showed in Mahadevan (1997).
The electron temperature $T_e$ is around $10^9$\,k, and is assumed
to be constant for $r<10^3$, where r is the dimensionless radius
of the accretion disk, defined in $R=rR_{Schw}=r\frac{2GM}{c^2}$.
The ion temperature given by Mahadevan (1997) is approximated to

\begin{equation} T_i=9.99\times10^{11}r^{-1} \ \mbox{k}
\end{equation}

The temperature profile is shown in Fig.~4. We simply assume an
optically thin accretion disk model and integrate the specific
emissivity over all radii to get the total luminosity. The ratio
between the luminosity of two-temperature and one-temperature
Bremsstrahlung radiation  $L_2/L_1$ is about 0.964 for
$T_e=10^9$\,k and 0.950 for $T_e=10^8$\,k, respectively. The small
difference between the two cases is due to the small difference
between the ion temperature and the electron temperature in most
of the accretion flow volume; only at small radii could $T_Z/T_e$
exceed 1000 where the two types of Bremsstrahlung radiation become
significantly different. We therefore conclude that in the ADAF
model the
correction due to the two-temperature Bremsstrahlung is negligible.\\

\begin{figure}[h]
  \begin{minipage}[t]{0.5\linewidth}
  \centering
  \includegraphics[width=65mm,height=50mm]{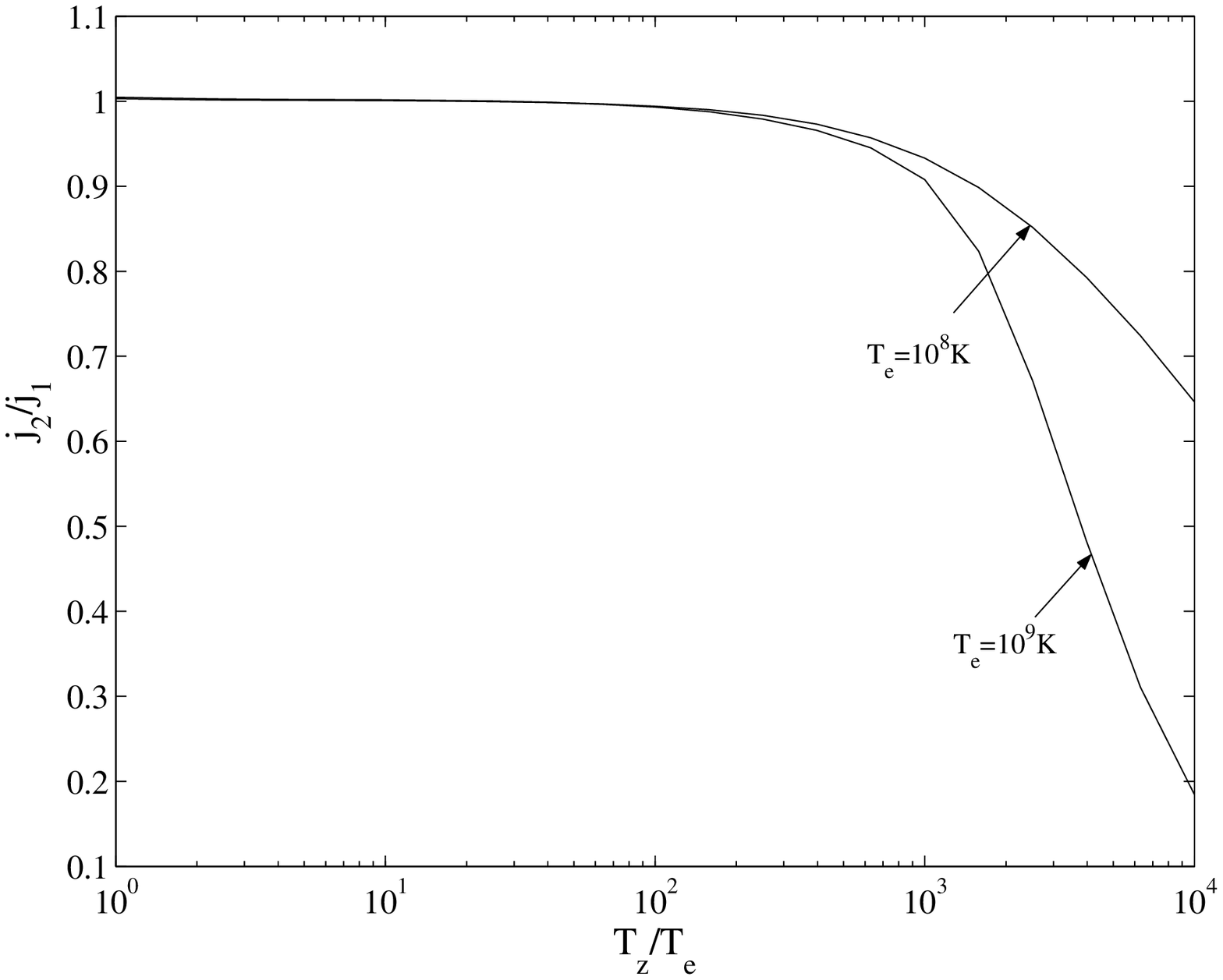}
  \vspace{-5mm}
  \caption{The ratio between the two
types of radiation emissivity ($j_2/j_1$) as a function of the
ratio between the ion temperature to the electron temperature
($T_Z/T_e$). If $T_Z/T_e$ is less than 100, $j_2/j_1$ is very
close to unity. When $T_Z/T_e$ is more than 1000, the difference
between $j_2$ and $j_1$ becomes significant.}
  \end{minipage}%
  \begin{minipage}[t]{0.5\textwidth}
  \centering
  \includegraphics[width=65mm,height=50mm]{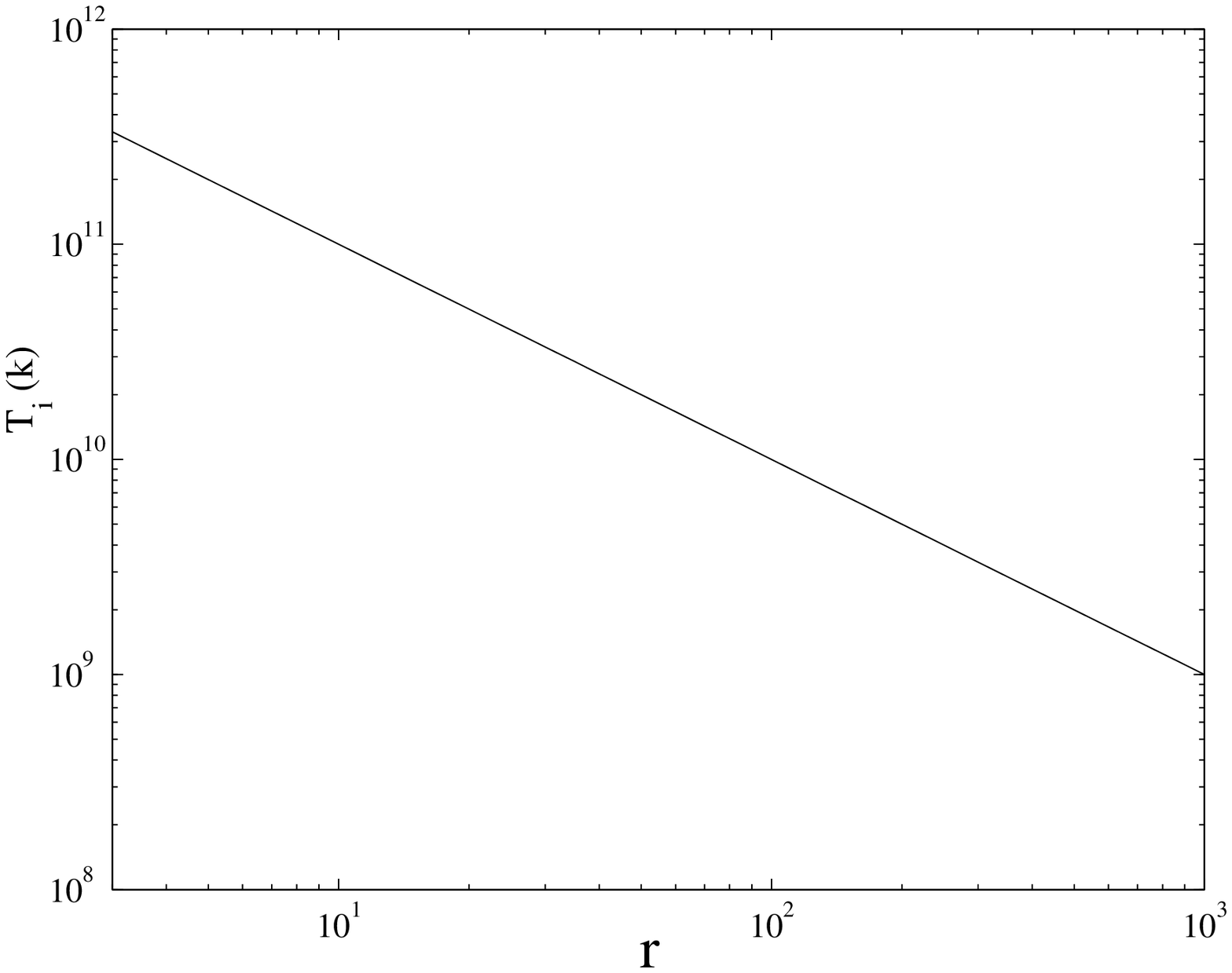}
  \vspace{-5mm}
  \caption{The ion temperature profile. The horizontal axis
  denotes the dimensionless radius $r$ ($R=rR_{Schw}=r\frac{2GM}{c^2}$) of the accretion disk, ranging from 3 to 1000.}
  \end{minipage}%
  \label{Fig:fig34}
\end{figure}

\section{Discussion}
\label{sect:discussion}
In a plasma with a high electron temperature, the bremsstrahlung
from electron-positron ($e^+e^-$), electron-electron ($ee$),
positron-positron ($e^+e^+$) collisions may become important
(Svensson, 1982). In our non-relativistic case, $e^+e^-$ pair
creation and annihilation can be neglected, then we only need to
consider the $ee$ bremsstrahlung, in addition to the $e$-proton
bremsstrahlung we have calculated above. For $T_e<10^9$\,k, we
calculate the cooling rates of electron-ion bremsstrahlung
($q_{ei}$) and $ee$ bremsstrahlung ($q_{ee}$) according to
Svensson (1982) and Narayan \& Yi (1995), and get that
$q_{ee}/q_{ei} < 0.3$. Therefore the electron-ion
bremsstrahlung dominates the radiation power and the $ee$ bremsstrahlung can also be neglected in non-relativistic cases.\\

In summary, our results show that the two-temperature
Bremsstrahlung radiation power is significantly lower than the
one-temperature Bremsstrahlung radiation if the ion temperature is
more than $1000T_e$. Although the temperature difference in the
ADAF model could exceed this critical value, the luminosity
correction due to this effect is still negligible due to the
rapid decrease of the ion temperature at large radii. However if in some more extreme
astrophysical environment the ion temperature is significantly higher than the
electron temperature, the two-temperature Bremsstrahlung radiation calculated in this work should be taken into account.\\

\begin{acknowledgements}
We thank the anonymous referees for valuable suggestions and
comments which have improved the manuscript significantly. This
study is supported in part by the Special Funds for Major State
Basic Research Projects (10233010) and by the National Natural
Science Foundation of China.
\end{acknowledgements}

\label{lastpage}

\end{document}